\documentclass[a4paper,11pt]{article}
\pdfoutput=1 

\usepackage{jinstpub} 
\usepackage{hyperref}
\usepackage{subcaption}
\usepackage{textcomp}
\usepackage{multirow}
\usepackage{multicol}\usepackage{array}
\usepackage{booktabs}

\usepackage{xcolor}
\usepackage{comment}
\usepackage[normalem]{ulem}

\usepackage[LGRgreek]{mathastext}

\usepackage{lineno}

\definecolor{darkgreen}{RGB}{0,128,0}
\newcommand\addedRT[1]{#1}
\newcommand\removedRT[1]{}

\title{\boldmath Radon Emanation from Dust of Varying Composition and Size}

\author[a,b,c,1]{Yue Meng,\note{Corresponding author.}}
\author[a]{Jerry Busenitz,}
\author[a]{Andreas Piepke,}
\author[a]{Raymond Tsang,}
\author[d]{Mengmeng Wu,}
\author[b]{Yukun Yao}

\affiliation[a]{Department of Physics and Astronomy, University of Alabama, Tuscaloosa, AL 35487, US}
\affiliation[b]{School of Physics and Astronomy, Shanghai Jiao Tong University, MOE Key Laboratory for Particle Astrophysics and Cosmology, Shanghai Key Laboratory for Particle Physics and Cosmology, Shanghai 200240, China}
\affiliation[c]{Shanghai Jiao Tong University Sichuan Research Institute, Chengdu 610213, China}
\affiliation[d]{School of Physics, Sun Yat-Sen University, Guangzhou 510275, China}

\emailAdd{mengyue@sjtu.edu.cn}

\abstract{
$^{222}Rn$ emanating from environmental dust constitutes an important background component for many low-energy, low-rate experiments. 
Radon emanation rates from dust and rock, thus, are important for experiment planning.  In this paper, we report measured radon emanation fractions for five types of dry dust differing in grain size and composition.  These data were obtained by a novel technique in $\gamma$-spectroscopy, measuring emanated and non-emanated $^{222}Rn$ progeny activities as well as the the parent $^{226}$Ra activity in a time series.   The range of observed radon emanation 
fractions is $(3.7 \pm 2.0)\%$ to $(16.2 \pm 0.9)\%$. Four of the five samples are standardized samples available commercially, so additional investigations of these samples may be readily carried out.}

\keywords{Radon Emanation, HPGe, Ultra-low Background Experiments}

\arxivnumber{2201.00699} 

\begin{document}
\maketitle
\flushbottom

\section{Introduction}
\label{sec:intro}
\noindent
$^{238}$U is present in rock typically at ppm-concentrations, together with its progeny, decaying at equal rates. 
Because of its relatively long half-life, the low-reactivity noble gas $^{222}$Rn (T$_{1/2}$=3.8235 d), one of the $^{238}$U progeny, may remain in or exit its host material by means of diffusion or recoil. This process of radon out-gassing is sometimes called emanation.
Because of its low reactivity, emanated $^{222}$Rn readily travels through gases and into experimental systems .
The $\beta$ and $\gamma$-radiation subsequently emitted in the decay of its daughters can then result in unwanted background signals. $\alpha$-emitting 
radon daughters can further contribute neutron-induced background events via nuclear $(\alpha,n)$-reactions 
on low-Z materials. Both 
sources of backgrounds -- direct and induced emissions -- 
constitute an important interference for rare-event searches such as dark matter and neutrinoless double-beta decay experiments~\cite{
Darwin_2016,
LUX:2016ggv,
XENON:2018voc,
EXO-200_final_2019,
cupid_pre-cdr_2019,
LZ:2019sgr,
finalcpc,
XENON:2020kmp,
CUPID_PRL_2021,
nEXO:2022nam,
legend_2021,
PandaX-4T:2021bab}.
More generally, due primarily to the risks to human health from environmental radon, radon emanation from dust, rocks, soil, and building materials has been the subject of many investigations; see ~\cite{amasi2015, kumarta2015,cameron1987,iaea2013} for examples and further references. Because of its relatively short half life of 55.6 s, the $^{232}$Th progeny $^{220}$Rn is rarely a background concern. It converts quickly into its reactive progeny, thus, becoming immobile. The study presented here focuses on $^{222}$Rn.

In soil, which typically exhibits U-chain secular equilibrium, the summed emanated and non-emanated  $^{222}$Rn activity equals that of its long-lived parent $^{226}$Ra (T$_{1/2}$=1600 y). The emanated $^{222}$Rn fraction can be measured by observing the in-growth of its daughters $^{214}$Pb and $^{214}$Bi. This can be done after removing the radon gas from a sample, allowing to observe how it grows back. Such removal leaves the non-emanated $^{222}$Rn fraction, and with the activities of its daughters, unchanged. 
It is known from the literature that the radon diffusion constants in intact dust grains are exceedingly small~\cite{nazaroff_nero_1988}, leading to diffusion lengths much shorter than typical ion recoil ranges. In such a situation $^{222}Rn$ release is mainly driven by nuclear recoil and one would not expect a strong temperature dependence. For the study presented here, all measurements were, therefore, performed at room temperature. The possible impact of moisture may warrant future studies of this point.

In this study, $\gamma$-spectroscopy is utilized to determine the decay rates of $^{226}$Ra, $^{214}$Pb and $^{214}$Bi. The $^{226}$Ra decay rate is expected to stay constant, serving as a convenient monitor. The observed time dependence of the $^{214}$Pb and $^{214}$Bi decay rates alone allows us to infer both the emanated and non-emanated radon fractions.
To demonstrate the temporal stability of the measurement system, the $^{226}$Ra decay rate was determined using the 186 keV $\gamma$-peak. The $^{214}$Pb and $^{214}$Bi decay rates, together with their time dependencies, were derived from their prominent $\gamma$-peaks at 295 keV and 352 keV ($^{214}$Pb), and 609 keV, 1120 keV and 1764 keV ($^{214}$Bi).\\

\section{Sample Characteristics}
Measurements with five different dust samples are reported in this paper, four of which were procured commercially\footnote{Powder Technology Inc., www.powdertechnologyinc.com}. These commercial samples came with certificates of analysis for grain size and composition. 
They were chosen to explore radon emanation for a range of compositions and grain sizes; further, for some of these samples, measurements may be found in the literature on grain shape and texture, which are useful for modeling. 
The choice of grain size distributions was guided by the availability of the commercial samples. The study presented here is limited to a phenomenological approach. While simple modeling has been attempted and is described below, an exhaustive study of all possible parameters impacting radon emanation was beyond the scope of this work.

The samples and their characteristics, such as average particle size and size distribution, are listed in Table \ref{tab:sampleinfo}.  
Samples A1 and A4, varieties of so-called Arizona test dust, have the same chemical composition but
exhibit different grain size distributions. Compared to samples A1 and A4, samples AFRL-02 and AFRL-03 have a different chemical composition. Compared to each other,
samples AFRL-02 and AFRL-03 have identical chemical compositions but show different median grain sizes of their quartz components.


The sample densities were estimated in our laboratory and served as input to the detector simulation code. The effective dust density relies on a geometrical estimation of the sample volume. We estimate the uncertainty of this quantity to be $\pm 10\%$. 
Note that the determination of the outgassing fractions from fits to the time dependence alone are independent of the detector efficiency modeling. 
The fifth sample is comprised of dust collected by a HEPA vacuum cleaner in the clean room at the Surface Assembly Laboratory (SAL) of the Sanford Underground Research Facility (SURF) and includes household waste. Larger debris was separated from the finer, more dust-like component, before performing measurements. 

\begin{table}[!th]
\begin{center}
\scalebox{0.6}{
\begin{tabular}{c|c|c|c|c}
\hline
Sample  & Composition & \begin{tabular}{@{}c@{}}Mass  \\$[g]$ \end{tabular} & \begin{tabular}{@{}c@{}}Average Density \\$[g/cm^3]$ \end{tabular}  & \begin{tabular}{@{}c@{}} Particle Size Average/Standard Deviation\\$[\mu m]$\end{tabular} \\ \hline
%
%
\multirow{4}[2]{*}{A1 Ultrafine} & 1-4\% Sodium oxide & \multirow{4}[2]{*}{103.1} &\multirow{4}[2]{*}{0.70}   & \multirow{4}[2]{*}{5.3/3.3 }\\
							     & 4-7\% Iron(III) oxide   &      & \\
							     & 1-2\% Magnesium oxide   &      & \\
							     & 0-1\% Titanium dioxide &
							     & \\ 
							   \cline{1-1}\cline{3-5}     
%
%
\multirow{4}[2]{*}{A4 Coarse} &  69-77\% Silica & \multirow{4}[2]{*}{124.1} & \multirow{4}[2]{*}{1.20}  & \multirow{4}[2]{*}{51/53} \\
							   & 8-14\% Aluminium oxide &      & \\
							   & 2.5-5.5\% Calcium oxide &       &\\ 
							   & 2-5\% Potassium oxide &        & \\
							     \hline
%
%
\multirow{3}[2]{*}{AFRL-02}  & 34\% Quartz  &  \multirow{3}[2]{*}{295.7} & \multirow{3}[2]{*}{0.59 }  & \multirow{3}[2]{*}{18/21} \\ 
					   &  30\% Gypsum &     &          \\
					   & 17\% Aplite &      &          \\ \cline{1-1}\cline{3-5} 
%
%
\multirow{2}[2]{*}{AFRL-03}  & 14\% Dolomite &  \multirow{2}[2]{*}{323.6} & \multirow{2}[2]{*}{0.65}  & \multirow{2}[2]{*}{26/25} \\  
					   & 5\% Salt &         &           \\ \hline				   
%
%
SURF SAL sweepings  
                        & unknown 
                        & \begin{tabular}{@{}c@{}} 163.4 (screened) \end{tabular}   
                        & 0.39   
                        & 328/242\\ 
                        \hline 
\end{tabular}
}
\end{center}
\caption{Characteristics of the dust samples used in this study: chemical composition, sample mass, density and average particle size. The average particle sizes and their variability were derived from the particle size distributions. The variability is, therefore, not a direct measure of uncertainty but a characteristic of the dust samples. Screened SURF SAL sweepings includes the dust component only.} 
\label{tab:sampleinfo}
\end{table}%

\begin{table}[!htbp]
\begin{center}
\scalebox{0.7}{
\begin{tabular}{c|c|c|c|c}
\hline
Sample & \% Mass Less Than & Quartz Grain Size [$\mu$m] & Gypsum Grain Size [$\mu$m] & Salt Grain Size [$\mu$m] \cr
\hline
\multirow{3}[2]{*}{AFRL-02}  & 10 & 1.0 $\pm$ 0.5 & 2.0 $\pm$ 1.0 & 1.0 $\pm$ 0.5 \cr 
 & 50 & 4.0 $\pm$ 1.5 & 11.5 $\pm$ 2.0  & 2.00 $\pm$ 0.75 \cr
  & 90 & 8.5 $\pm$ 2.5 & 38.0 $\pm$ 3.0 & 4.0 $\pm$ 1.0 \cr \hline
\multirow{3}[2]{*}{AFRL-03} & 10 & 5.0 $\pm$ 1.0 & 2.0 $\pm$ 1.0 & 1.0 $\pm$ 0.5 \cr 
 & 50 & 24.0 $\pm$ 2.5 & 11.5 $\pm$ 2.0  & 2.00 $\pm$ 0.75 \cr
  & 90 & 51.0 $\pm$ 4.5 & 38.0 $\pm$ 3.0 & 4.0 $\pm$ 1.0 \cr
  \hline
  \end{tabular}
  }
\end{center}
\caption{Grain size distributions for the quartz, gypsum, and salt components of the 
AFRL dust samples.  These components are blended with the dolomite (Dolocron 40-13) and aplite (Minspar 200) components after the dolomite and aplite have been sieved with a minus 200 mesh. Different components remain different grains.}
\label{tab:afrlgrain}
\end{table}

\begin{figure}[!h]
     \centering
     \begin{subfigure}[b]{0.45\textwidth}
         \centering
         \includegraphics[width=\textwidth]{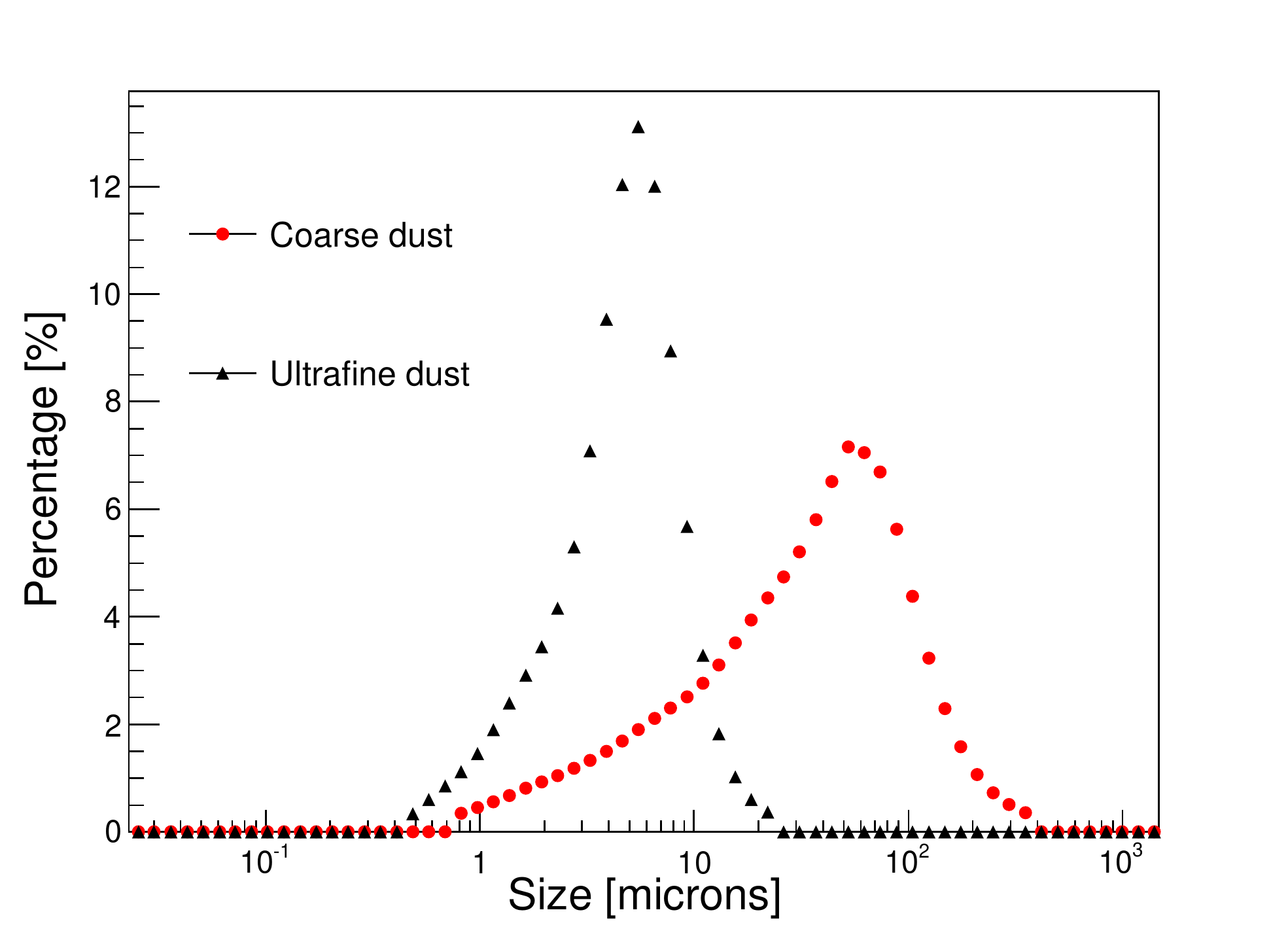}
         \caption{A4 Coarse and A1 Ultrafine Dust}
         \label{fig:sizea}
     \end{subfigure}
     \hfill
     \begin{subfigure}[b]{0.45\textwidth}
         \centering
         \includegraphics[width=\textwidth]{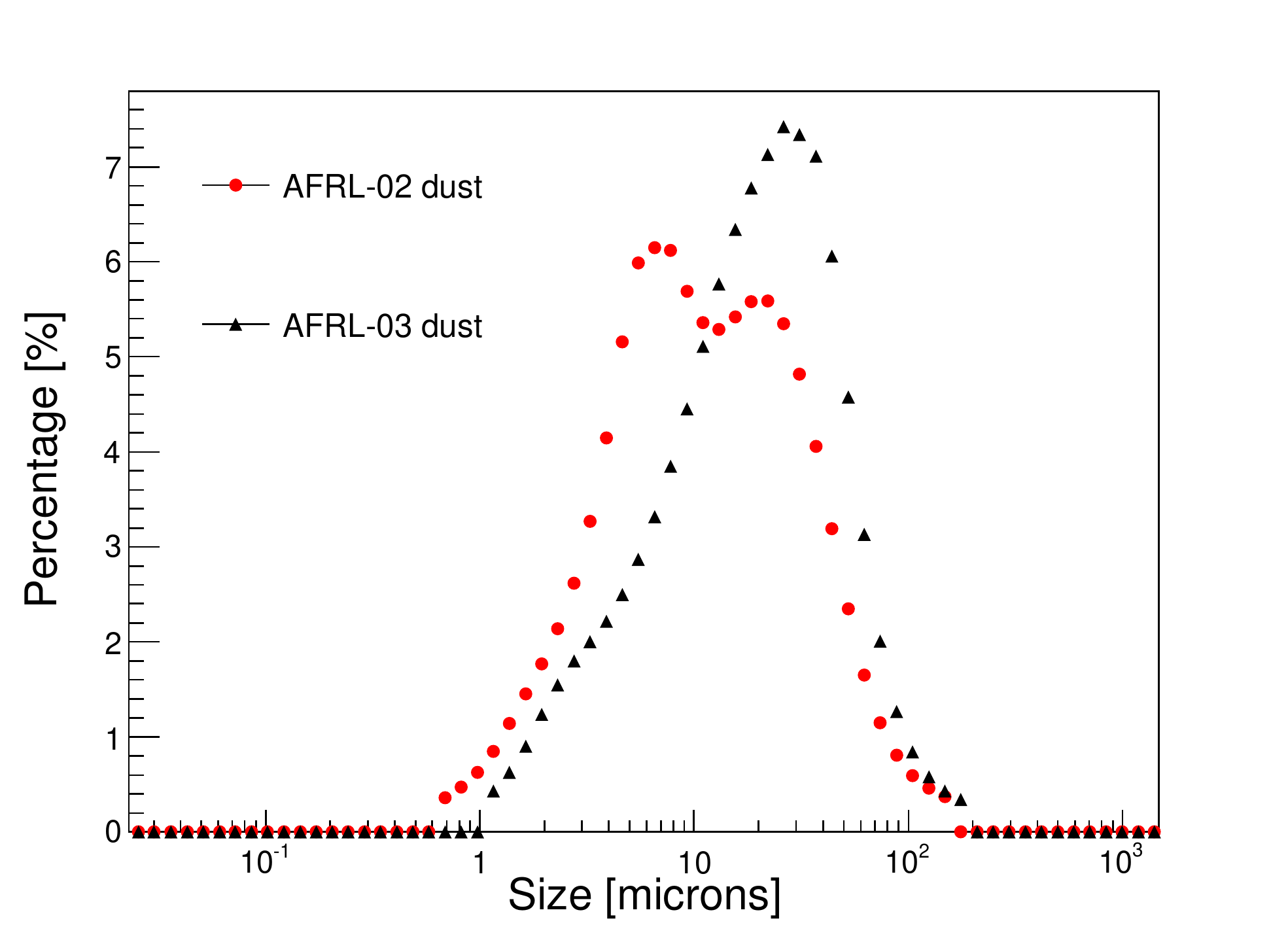}
         \caption{AFRL-02 and AFRL-03 Dust}
         \label{fig:sizeb}
     \end{subfigure}
     \hfill
     \begin{subfigure}[b]{0.45\textwidth}
         \centering
         \includegraphics[width=\textwidth]{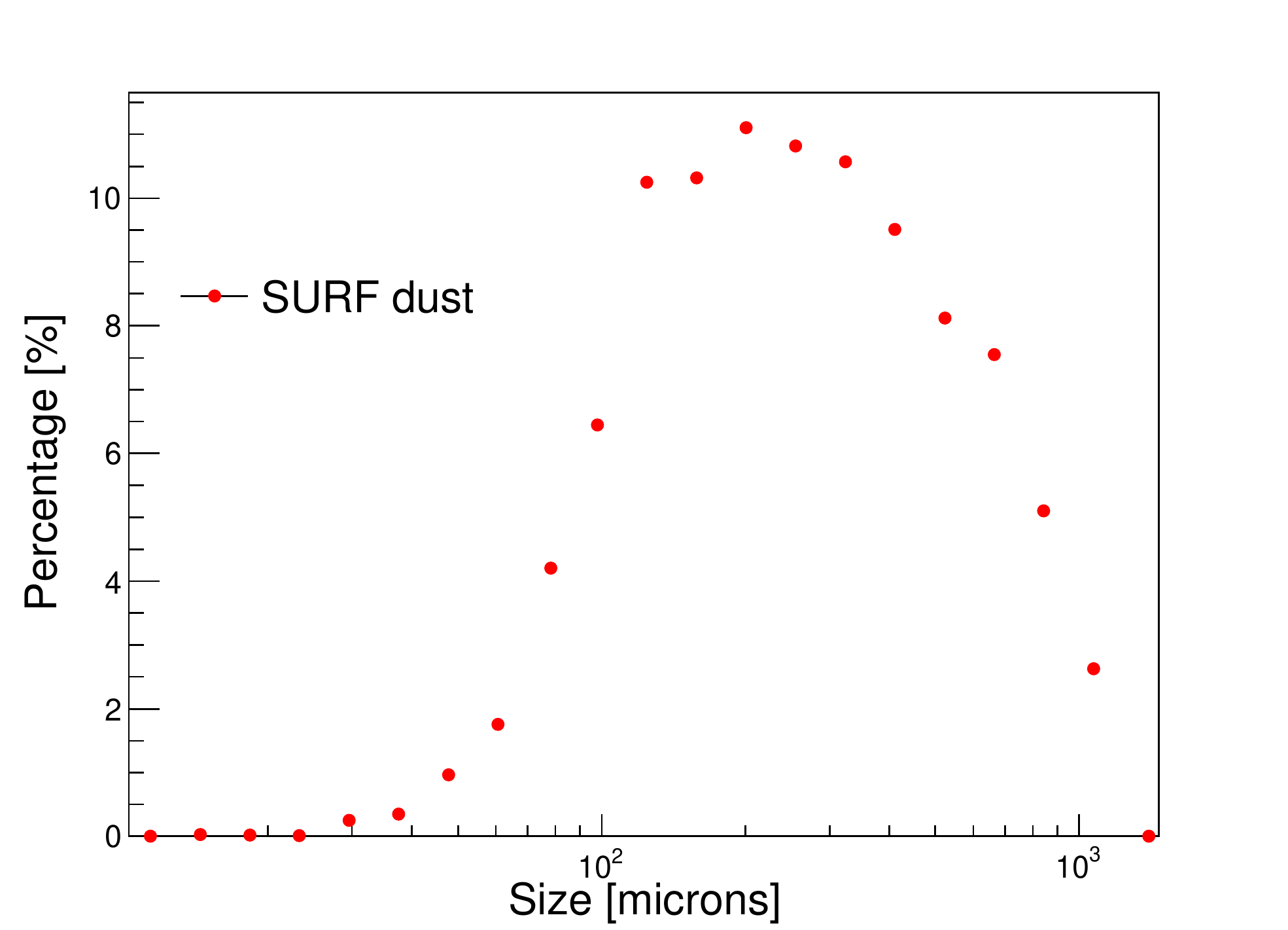}
         \caption{SURF SAL dust with 1000 $\mu$m sieve mesh filtration}
         \label{fig:sizec}
     \end{subfigure}
        \caption{Particle size (principal axis) distributions of dust samples}
        \label{fig:particlesize}
\end{figure}

%
The particle size distributions for the commercial samples, as supplied by the vendor, are shown in Figure \ref{fig:sizea} and \ref{fig:sizeb}. Before determining the particle size distribution of the SURF SAL dust sample, shown in Figure \ref{fig:sizec}, it was filtered with a 1000 $\mu$m mesh sieve to separate debris from dust. The sample was then measured by Bettersize Instruments Ltd. in China\footnote{Bettersize Instruments Ltd, https://www.bettersizeinstruments.com}.
As a further cross check of the size assessment, samples of A1 and A4 dust were submitted to Bettersize Instruments Ltd. for analysis. The size distributions obtained by them were found to be in qualitative agreement with the vendor data.

Concerning the shape and texture of the grains, measurements have been reported for Arizona test dust samples derived from the same feed stock and overlapping in grain size with the Arizona test dust samples investigated in this work. These studies indicate that grain shape and texture are independent of grain size and grain shape deviates significantly from spherical.  Details can be found in references~\cite{fletcher2000,woodward2015,wagner2008,connolly2020}. 
\color{black}

\section{Procedure}
To investigate a possible relation between the radon emanation fraction and average particle size or composition, the following measurement procedure was followed for samples.  The sample was spread out in a thin layer, a few mm thick, facilitating ventilation of radon. The sample was ventilated in a fume hood with ambient air for a few days.  All emanated radon is presumed to escape in this process.   The sample was then tightly wrapped and sealed into either a Mylar bag of 0.0762 mm thickness (UA) or double Nylon bags of 0.1 mm thickness (Jinping).
%
%
Mylar and Nylon bags are excellent radon barriers~\cite{radonpermeability}, prohibiting radon atoms 
to escape to any significant degree. Without further delay, the sealed sample was placed in one of two  shielded low-background HPGe detector setups utilized in this study. Samples were counted for a period of 10--14 days. The impact of varying the water content of the A4 sample was found to be small, as shown in Table~\ref{tab:result}. A larger water content could well result in a larger impact.\\

The dust samples, as received for our measurements, were dry.  During the first step of the procedure, however, in which the radon was vented, the dust was exposed to laboratory air in which the relative humidity varied in the range 30-50\%. To investigate whether this introduced a systematic effect in our measurement results, the first sample was dried by pumping and baking for 24 h at 200\textdegree{} C. The sample was sealed into double 0.1 mm Nylon bags. 
A second sample of A4 dust was exposed to moist air (relative humidity $\sim$70\%) for 24 h and also sealed in double Nylon bags. Both samples were then counted for $\sim$10 days. \\

Sample counting utilized two different shielded low background Ge detector setups.
\begin{itemize}
    \item Most sample counting was performed with the above-ground GeII setup, operated at the University of Alabama (UA). The setup consist of a low background Canberra p-type coaxial germanium detector with copper and lead shielding. Large Bicron plastic scintillation panels are utilized as active muon veto system. 
    \item The de-watering studies, as well as comparative measurements with the A4 sample, were performed with the Jinping
    (JP) Ge2 setup, located in the Jinping underground laboratory in China. The setup utilizes a low background Canberra germanium detector and is equipped with a copper and lead passive radiation shield. 2400 m of rock provide cosmic rays shielding.
\end{itemize}

\begin{figure}[!h]
\centering
\begin{subfigure}{.4\textwidth}
  \centering
  \includegraphics[width=0.8\linewidth]{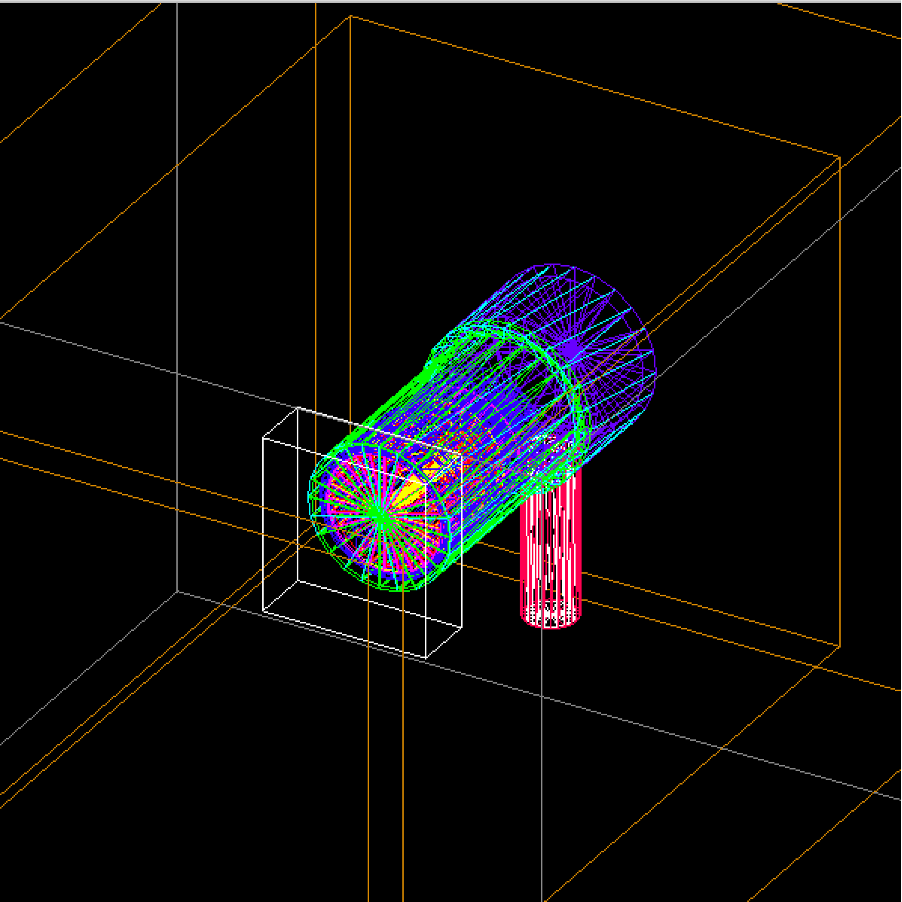}
\end{subfigure}%
\begin{subfigure}{.6\textwidth}
  \centering
  \includegraphics[width=1.0\linewidth]{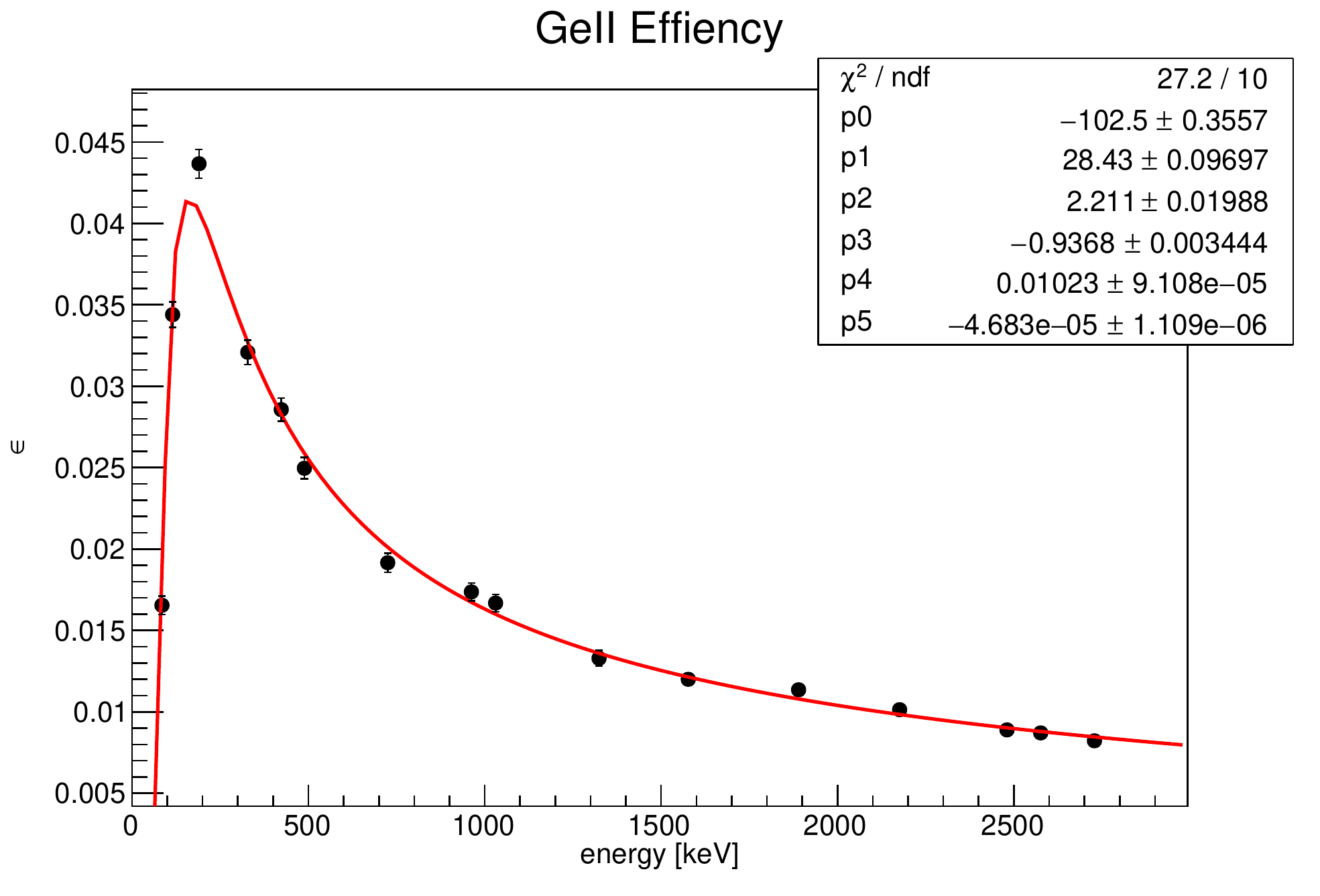}
\end{subfigure}
\caption{GEANT4 visualization of the GeII sample geometry and HPGe detection efficiency $\varepsilon$  vs. energy $E$ parameterization: $\varepsilon(E)= p0 + p1\cdot log(E) + p2\cdot log^2(E) + p3\cdot log^3(E) + p4\cdot log^5(E) + p5\cdot log^7(E)$ }
\label{fig:simulation}
\end{figure}%

Counting rates were converted into nuclide-specific decay rates using energy-dependent detection efficiencies, determined by means of GEANT4 detector simulations, as described in~\cite{Tsang:2019apx}. The left panel of Figure \ref{fig:simulation} shows a rendering of the simulated geometry for the UA GeII setup.
An example for an efficiency curve and a parametric fit to the Monte Carlo data is shown in the right panel of Figure \ref{fig:simulation}. This conversion does not directly impact the measurement of the outgassing fraction.

\begin{figure}[!th]
\begin{center}
\includegraphics[width=13cm]{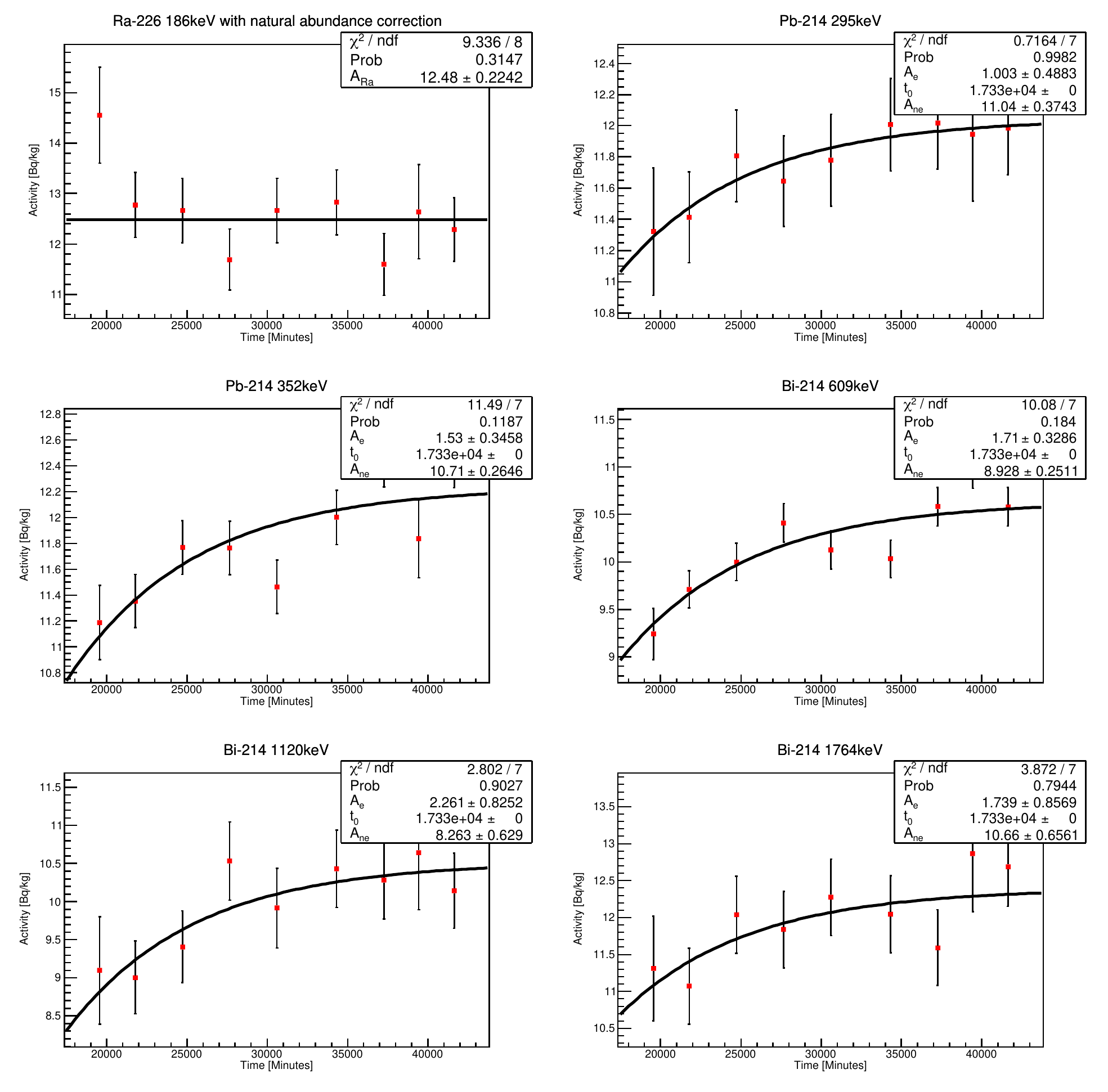}\\
\caption{In-growth of emanated radon daughters observed for the dust sample from SURF SAL sweepings. The parameter $t_0$ stated in the fit window denotes the reference time-zero since sealing of the bags. The time constant used in the fits has been fixed to the tabulated $^{222}$Rn half life.}
\label{fig:6plots}
\end{center}
\end{figure}

An example of the observed in-growth of radon progeny and of the time fits used to determine the transient and constant activity fractions is shown in Figure~\ref{fig:6plots} for the SURF SAL sample. 
$^{226}$Ra was detected via the 186 keV peak, $^{214}$Pb via the 295 keV and 352 keV lines, and $^{214}$Bi via the 609 keV, 1120 keV, and 1764 keV peaks. $\gamma$-radiation with 186.2 and 185.7 keV is emitted in the decay of $^{226}$Ra ($^{238}$U series) and $^{235}$U, respectively. The resulting full absorption peaks cannot be resolved from each other. The determination of the $^{226}$Ra activity accounts for this degeneracy, assuming $^{235}$U and $^{238}$U are present in the samples with their natural isotopic abundances. 
\color{black}
The solid lines in Figure~\ref{fig:6plots} show the time functions fit to the data. 
The constant, $A_{ne}$, and transient, $A_{e}$, $^{214}$Pb and $^{214}$Bi activity terms were determined by fitting the data with:
\begin{equation}
    A(t)\; =\; A_{ne}+A_e\cdot \left(1-e^{-t/\tau_{Rn}} \right)
\end{equation}
with the mean life time fixed to that of $^{222}Rn$ ($\tau_{Rn}=7943.2\; min$).
$A_{e}$ and $A_{ne}$ are interpreted as the emanated and non-emanated $^{222}Rn$ progeny fractions.
Assuming secular equilibrium and lossless radon collection, $A_{Rn}=A_{e}+A_{ne}$ is equal to the $^{222}Rn$ decay rate.
For reporting purposes, the various peak activities were combined by means of a weighted average. The peak-wise radon outgassing fractions is determined as: 
\begin{equation*}
    f_i=\frac{A_{e,i}}{A_{e,i}+A_{ne,i}},
\end{equation*}
where the index $i$ labels the various $\gamma$-peaks listed above. 
Note that the Ge detector efficiency cancels in this ratio. The reported outgassing fractions are, thus, independent of the Monte Carlo simulation and its uncertainty. The outgassing fraction of each sample was then determined as the weighted mean of the appropriate $f_i$-values. The error determination for the $f_i$-values accounts for the strong anti-correlation observed for the uncertainties of $A_{e}$ and $A_{ne}$.
Calculating the emanation fractions from averaged activities instead leads to very similar results.
\color{black}

\begin{figure}[!th]
\begin{center}
\includegraphics[width=7cm]{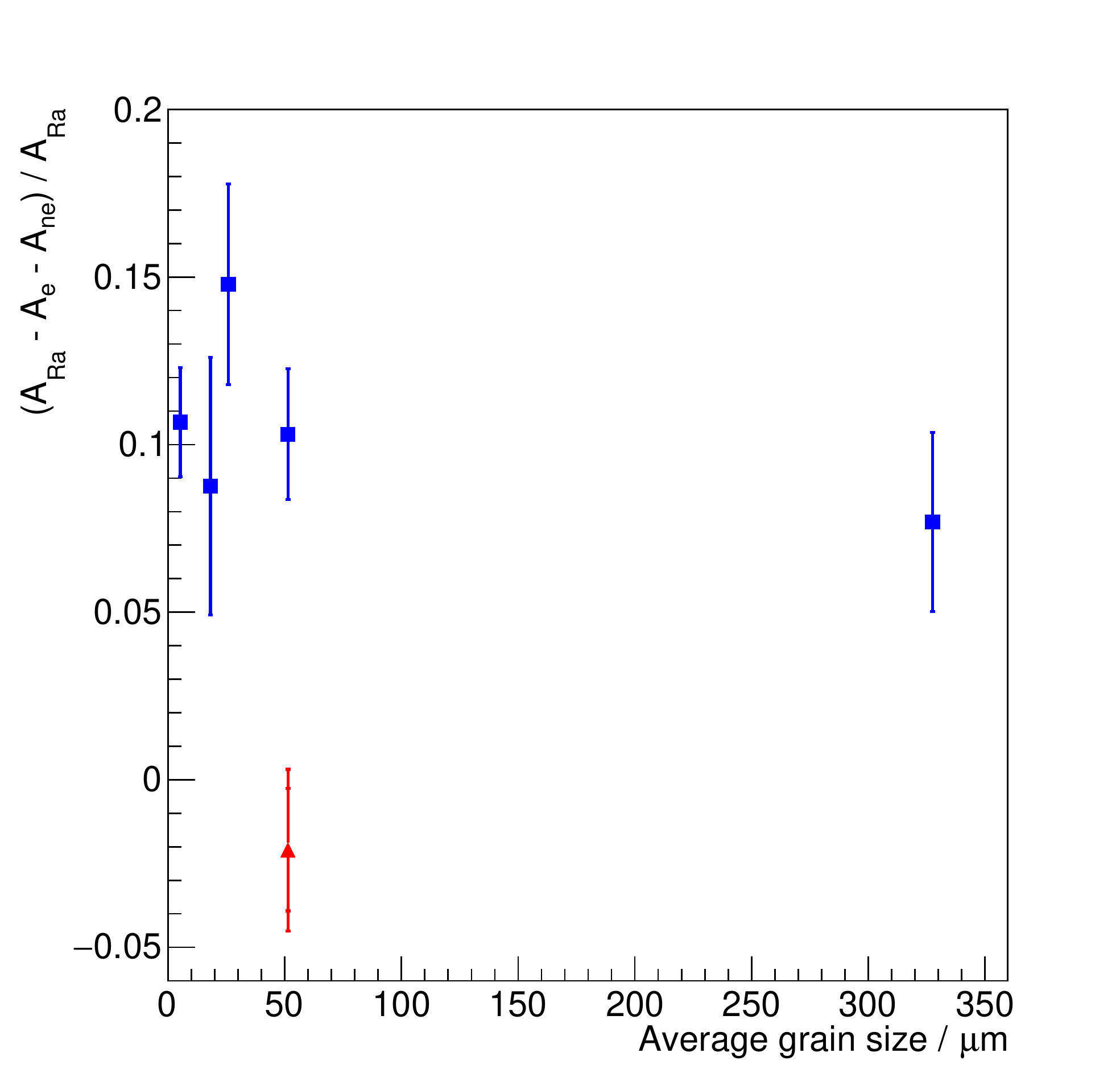}\\
\caption{Relative difference $\Delta$ of the $^{226}$Ra and $^{222}$Rn activities as a function of average grain size. The blue squares represent data taken at UA, while the red triangles represent data taken at Jinping.}
\label{fig:rel_diff}
\end{center}
\end{figure}

Note in Figure~\ref{fig:6plots} that the $^{226}$Ra activity, $A_{Ra}$, was found to be constant, as expected. The growth of the $^{214}$Pb and $^{214}$Bi activities follows the $^{222}$Rn mean life time, also as expected. The other sample data behaves in a similar way.
%
%
In order to study the $^{226}$Ra-$^{222}$Rn balance, Figure~\ref{fig:rel_diff} shows the relative activity difference $\Delta=\frac{A_{Ra}-A_e-A_{ne}}{A_{Ra}}$ for the different dust sample measurements. 
%
%

One would expect $\Delta$ to be compatible with zero in case all $^{222}$Rn production and disappearance has been accounted for. The high points in Figure~\ref{fig:rel_diff} correspond to measurements made with detector UA GeII, and the two low points to those made with JP Ge2. The plot shows a 10\% upward bias for the UA GeII data and a 2\% downward bias for the JP Ge2 points, relative to the average. We interpret this data to show that the radon production and disappearance is accounted for within 10\% uncertainty. The average density of the dust and approximate bag sizes were used for the simulation of two HPGe detectors. This is an additional source of uncertainty. 
As pointed out above, the outgassing fractions reported here do not directly depend on the $^{226}$Ra activities, measured directly via the 186 keV full absorption peak. Figure~\ref{fig:rel_diff} serves to demonstrate that the observed time dependence is not an instrumental artefact. It further shows that the relevant radon progeny activities are accounted for to within 10\%, a value compatible with our estimate of the systematic uncertainty of the detector efficiency estimation.

\section{Results and Discussion}

\begin{figure}[!th]
\begin{center}
\includegraphics[width=9cm]{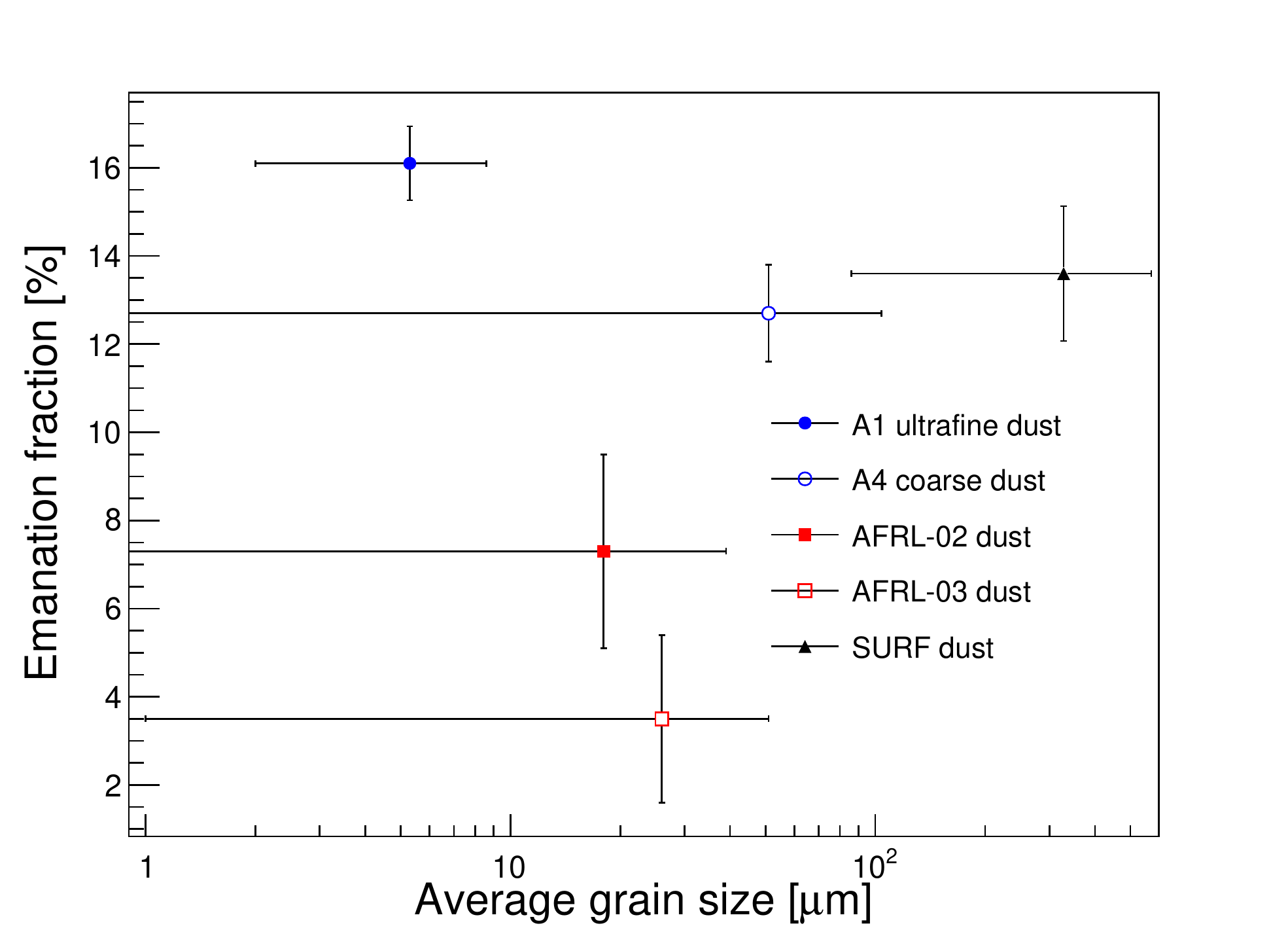}\\
\caption{Observed emanation fractions versus average particle size. Blue points correspond to the A-type dust samples, points in red to AFRL-dust. The black point denotes the SURF SAL sample. The vertical error bars correspond to the measurement uncertainty of the emanation fraction. The horizontal error bars indicate the standard deviations of the respective particle size distributions. The A4 data point corresponds to the weighted average of all measurements reported in Table~\ref{tab:result}.}
\label{fig:eman_frac}
\end{center}
\end{figure}

The measured $^{226}Ra$ activities, the averaged emanated and non-emanated activities,  and average emanation fractions are given in Table~\ref{tab:result} for all samples. All averages are variance weighted and performed peak-wise for each sample.  Figure~\ref{fig:eman_frac} depicts the observed radon emanation fraction plotted versus the average particle size.   
%
%
The radon emanation fractions from dust observed in this study lie between $(3.7 \pm 2.0)\%$ and $(16.2 \pm 0.9\%)$, as shown in Table~\ref{tab:result}.

\begin{table}[!th]
\begin{center}
\scalebox{0.7}{
\begin{tabular}{c|c|c|c|c}
\hline
Sample  & \begin{tabular}{@{}c@{}}$^{226}Ra$ activity \\$[Bq/kg]$ \end{tabular}& \begin{tabular}{@{}c@{}}Emanated \\$^{214}Pb-^{214}Bi$ activity  \\$[Bq/kg]$ \end{tabular}  & 
\begin{tabular}{@{}c@{}} Non-emanated \\$^{214}Pb-^{214}Bi$ activity \\$[Bq/kg]$\end{tabular}  &
Emanation fraction \addedRT{[\%]}\\ \hline \hline
\begin{tabular}{@{}c@{}}A1 Dust  \\($5.3\pm 3.3\; \mu m$) \end{tabular} &  
46.9 $\pm$ 0.58  & 
6.8 $\pm$ 0.41  & 
35.1 $\pm$ 0.27 & 
$16.2 \pm 0.9$\removedRT{\%} \\ \hline
\begin{tabular}{@{}c@{}}A4 Dust \\($51.4\pm 52.5\; \mu m$) \end{tabular} & 
29.1 $\pm$ 0.43  & 
3.3 $\pm$ 0.31  & 
22.8 $\pm$ 0.20 & 
$12.6 \pm 1.1$\removedRT{\%} \\ \hline
\begin{tabular}{@{}c@{}}AFRL-02 Dust \\($18.2\pm 20.7\; \mu m$)  \end{tabular}  & 
4.9 $\pm$ 0.13   &
0.33$\pm$ 0.11    & 
4.2 $\pm$ 0.08 & 
$7.4 \pm 2.3$\removedRT{\%}\\ \hline
\begin{tabular}{@{}c@{}}AFRL-03 Dust  \\($25.9\pm 24.5\; \mu m$) \end{tabular}  &  
6.6 $\pm$ 0.14  & 
0.2 $\pm$ 0.11  & 
5.4 $\pm$ 0.08 &  
$3.7 \pm 2.0$\removedRT{\%}  \\ \hline \hline
\begin{tabular}{@{}c@{}}SURF SAL sweepings  \\($327.5\pm 242.2\ \mu m$) \end{tabular}  & 
12.5 $\pm$ 0.22    & 
1.6 $\pm$ 0.20   & 
10.0 $\pm$ 0.15 & 
$13.5 \pm 1.7$\removedRT{\%}   \\ \hline \hline
\begin{tabular}{@{}c@{}}A4 Dust \\(Dry sample) \end{tabular}  & 
24.0 $\pm$ 0.37  & 
2.7 $\pm$ 0.19  & 
21.8 $\pm$ 0.14 & 
11.0 $\pm$ 0.8\removedRT{\%} \\ \hline
\begin{tabular}{@{}c@{}}A4 Dust \\(Moist sample) \end{tabular}  & 
23.8 $\pm$ 0.38  & 
2.2 $\pm$ 0.37  & 
22.1 $\pm$ 0.22 & 
$9.2 \pm 1.5$\removedRT{\%} \\ \hline
\end{tabular}
}
\end{center}
\caption{$^{226}Ra$ activities, derived from the 186 keV peak, were corrected for the natural abundances of $^{235}$U and $^{238}$U. Statistical uncertainties are listed. A 10\%/15\% systematic uncertainty is assigned to the efficiency corrections of UA Ge2/JP Ge2.}
\label{tab:result}
\end{table}%

The radon emanation fractions of the samples we have measured are consistent with the fractions observed for similar materials~\cite{sakoda2011}.  According to the analysis of a comprehensive compilation of emanation fraction measurements carried out in Reference \cite{sakoda2011}, 0.03 is a typical value for mineral dusts while typical emanation fractions for rocks and soils are 0.13 and 0.20, respectively.  The AFRL samples are primarily a mixture of a few minerals.   The feed stock of the A1 and A4 samples is naturally-occurring so-called Arizona dust, which is presumed to be comprised of fine particles of local soil and rock. The SURF Surface Assembly Lab (SAL) sample would also seem likely to be small bits of soil and rock local to the Sanford Underground Laboratory.     

Our measurements were carried out under normal laboratory conditions without tight control on temperature or humidity. The check on possible effects of humidity variations--where we repeated one measurement under very humid conditions--indicates that humidity variations did not affect our results.

Based on published measurements of various grades of Arizona test dust summarized in the section on sample characteristics, the A1 and A4 samples may be taken as practically identical except for grain size distributions.  According to our measurements, the A1 sample with the smaller grain size exhibits a significantly larger emanation fraction.  The AFRL samples are known to be identical in composition and also have the same grain size distributions except the quartz grains are rather smaller, on average in the AFRL-02 sample than in the AFRL-03 sample.  The AFRL-02 sample with the smaller quartz grains is observed to have a higher emanation fraction but with limited statistical significance. Note also that the AFRL samples may differ in other ways, in grain shape and texture, for example, so whatever difference in emanation fractions there could be due to several factors, not just grain size. In order to evaluate the effect of grain shapes on radon emanation fractions, we performed calculations 
of the recoil-driven ejection of $^{222}$Rn ions from single grains, using GEANT4, comparing spherical and oblately spheroidal grain shapes. As shown in reference \cite{woodward2015}, the grains have oblate spheroidal shape with an aspect ratio of 8. The result shows that the radon emanation fraction is a factor of 4 larger for oblate spheroids having an equatorial diameter eight times greater than the polar diameter. This demonstrates the importance of the knowledge on the shape of the grains. This simple recoil modeling results in outgassing fractions a factor of 6-22 smaller than observation for the commercial samples, assuming spherical dust particles. 
The value obtained for the SURF dust in this way is a factor of 430 smaller than observation for the spherical model. For more complicated models, factors influencing radon emanation include how radium is distributed throughout the grain, whether the grains are isolated (single grain) or reside in bulk (multigrain), the shape and texture of the grains, water content, and temperature \cite{sakoda2017,SAKODA2010204,STAJIC201419,Chitra}.

Given the complexity of factors affecting radon emanation, it is clear that accurate modeling benefits from detailed information about the samples.   The advantage of measuring standardized materials such as A1, A4, AFRL-02, and AFRL-03 is that it is possible to procure additional samples for studies to fill in missing information on properties and to investigate how the emanation fraction may depend on measurement and environmental conditions.  An alternate approach is to collect a sample of dust which could contaminate a detector and directly measure radon emanation from it.  This was the motivation for measuring the SURF SAL sweepings sample, taken from the lab in which the LZ detector was being assembled. 
LZ is a search for elastic scattering of dark matter particles off xenon nuclei, described in reference~\cite{LZ:2019sgr}.
The observed emanation fraction is similar to what was observed for the Arizona test dust samples and much larger than what was observed for the AFRL samples.

\section{Acknowledgements}
This work was orignally conceived as a contribution to the construction of the LZ experiment. We thank our LZ colleagues for their support. We are especially grateful to Kevin Lesko  and the late Al Smith of Lawrence Berkeley National Laboratory (LBNL) for bringing the radon progeny time method to our attention. We thank Bettersize Instruments Ltd to their help in measuring the particle size distributions.  This research was supported in part by the U.S. Department of Energy under DOE Grant DE-SC0012447, by a grant from the National Science Foundation of China (No. 12005131), Shanghai Pujiang Program (No.19PJ1405800) and a grant from Sichuan Science and Technology Program (No.2020YFSY0057).

%

\end{document}